# Text Independent Speaker Identification System for Access Control

Oluyemi E. Adetoyi
Dept. of Electrical and Electronic Engineering
University of Ibadan
Ibadan, Nigeria
oe.adetoyi@ui.edu.ng

***Abstract*— Even human intelligence system fails to offer 100% accuracy in identifying speeches from a specific individual. Machine intelligence is trying to mimic humans in speaker identification problems through various approaches to speech feature extraction and speech modeling techniques. This paper presents a text-independent speaker identification system that employs Mel Frequency Cepstral Coefficients (MFCC) for feature extraction and k-Nearest Neighbor (kNN) for classification. The maximum cross validation accuracy obtained was 60%. This will be improved upon in subsequent research.**

## I. INTRODUCTION

Speaker recognition is a behavioral biometric system that is used to verify a person claimed identity from the spoken words. The system can be used in identification mode, whereby the system decides if the speaker is among a group of persons or in verification mode that determines who the speaker is [1]. The main goal in speech recognition is to convert the speech waveform into a set of features or vectors at a considerably lower information rate for further analysis. The feature extraction is usually performed in three stages. The first stage performs spectra temporal analysis of the signal and generates raw features describing the envelope of the power spectrum of short speech intervals. The second stage compiles an extended feature vector composed of static and dynamic features. While the last stage, which is not often present, transforms these extended feature vectors into more compact and robust vectors that can be supplied to the recognizer.

While there is no particular feature suitable for specific application, the choice of features should however allow automatic discrimination between different similar sounding speech sounds. Also, creation of acoustic models for these sounds without the need for an excessive amount of training data should be possible, and the statistics exhibited should be largely invariant across speakers and speaking environment. Obtaining statistically relevant information from incoming data, often requires mechanisms for reducing the information of each segment in the audio signal into a relatively small number of parameters, or features that describe each segment in characteristic way that other similar segments can be grouped together by comparing their features. Linear prediction Coefficients (LPC), Mel Frequency Cepstral Coefficients (MFCC) and Perceptual Linear Prediction (PLP) have been identified as the most well-defined acoustic feature extraction approaches [2]. Other feature extraction techniques can be found in [3] - [4].

Several models have been applied to perform feature matching, which entails estimation and comparison of the testing sample with the training sample. Some of these are: Gaussian Mixture Model (GMM), Gaussian Mixture Model based Universal Background Model (GMM-UBM), Vector Quantization (VQ), Hidden Markov Models (HMMs), Support vector Machines (SVM) and Artificial neural Networks (ANN) among others [2], [5].

This work proposed a text independent speaker identification system that utilize MFCC for feature extraction and K-Nearest Neighbor (KNN) for classification. In section two of the paper, similar works were discussed, while section three discussed the design methodology of the proposed system. Results obtained were presented in section four and section five provides the conclusion.

## II. RELATED WORKS

The speaker identification system in [2] makes use of MFCC and Normalized Pitch Frequency (NPF) to extract speech features, while ANN was utilized for classification. The use of NPF enhances the performance of the system over using MFCC alone. The use of Discrete Cosine Transform (DCT) and wavelet denoising pre-processing stages are also found to improve the system. However, the validation and prediction accuracy are not known.

The performance of MFCC and Power Normalized Cepstral Coefficients (PNCC) were compared in [1] for text independent speaker identification system using 16 coefficients each. Effect of linear channel was mitigated using Cepstral Mean-Variance Normalization (CMVN) and feature warping respectively. Eight different speakers from Grid-Audiovisual database, comprising of two females and six males were modelled using GMM-UBM. Both MFCC and PNCC have identification accuracy of 100%, but PNCC has higher score for ability to identify female than MFCC. Result also shows that feature warping achieved better results for both PNCC and MFCC than CMVN.

A multiscale chaotic feature extraction approach was proposed in [6] for speaker recognition. A multiresolution analysis technique was used to capture finer information on different speakers in the frequency domain. The speech chaotic characteristics based on the nonlinear dynamic model was extracted to improve the discrimination of features. A GMM-UBM was then used for classification. The Equal Error Rate (EER) under clean and noisy speech condition was reduced in comparison to acoustic feature, nonlinear feature and i-vector feature. However, the feature extraction technique is not optimized.

A hybrid speaker identification model, that incorporates pitch frequency coefficient from speech time domain analysis to enhance accuracy of features extracted using Mel frequency spectrum coefficients (MFCC) was presented in [7]. A single hidden layer feed-forward neural network (FFNN) tuned by an optimized particle swarm optimization (OPSO) algorithm was deployed for classification. The model was tested using 10-fold cross-validation over different levels of Adaptive White Gaussian Noise (AWGN) (0-50 dB). A recognition accuracy of 97.83% was obtained from the proposed model in clean voice environments. A noisy channel has lesser impact on the proposed model as compared with other baseline classifiers such as plain-FFNN, random forest (RF), K-nearest neighbor (KNN), and support vector machine (SVM).

MFCC was used to extract features from PhotoPlethysmoGraphy (PPG) of patients, while ANN was employed for classification by [8]. An accuracy of 100% and 98.07% was obtained for hold-out and 10-fold cross validation respectively for a dataset of 35 healthy patients. However, the accuracy of unhealthy patients was not taking into consideration.

The subtask of understanding speeches through the detection of emotions elicited by speakers while talking was the subject of research in [9]. The features of audio recordings from Ryerson Audio-Visual Database of Emotional Speech and Song (RAVDESS) was extracted using MFCC. The Convolutional Neural Network model was trained to classify six Ekman emotions (happy, sad, angry, fearful, disgust and surprise) in addition to calm and neutral emotions. Angry has the highest F1 score of 0.95 and sad has the lowest with 0.87, the mean F1 score is 0.91.

The task of converting Urdu spoken language into computer readable text was undertaken in [10]. The acoustic features of Urdu corpus were extracted using MFCC. Then the Cepstral Mean and Variance Normalization (CMVN) of the features were computed. These statistical models: Monophone, Tri-1, Tri-2, Tri-3 and Subspace Gaussian Mixture Model (SGMM2) are trained on Kaldi platform using 100 hours of Urdu language transcribed data. The subspace Gaussian model gives a word error rate of 9%.

## III. PROPOSED SPEAKER IDENTIFICATION SYSTEM

The block diagram of the proposed system is as shown in Fig. 1. In the training or enrolment phase, the features of acquired speech data are extracted using MFCC. The extracted features were then used to train the KNN model of the speaker identification system. During the test phase, the features extracted from untrained speech dataset was used by the kNN to predict the corresponding class of input data. Access can then be granted to the speaker if the prediction is true. If the prediction is false, the speaker will be required to provide password. A match password will thus confirm identity and access will be granted, otherwise it will be denied. The system was implemented in MATLAB environment.

### A. Data Acquisition

The dataset comprises of 1500 audio clips from speech made by five notable Presidents, making a total of 7500 audio clips. The Presidents are a mix of two women and two men. Some audio clips contain no speech, except applause of audience or other background noise. Some audio clips have large occupancy of leading or trailing noise, while the speech in some clips are not audible.

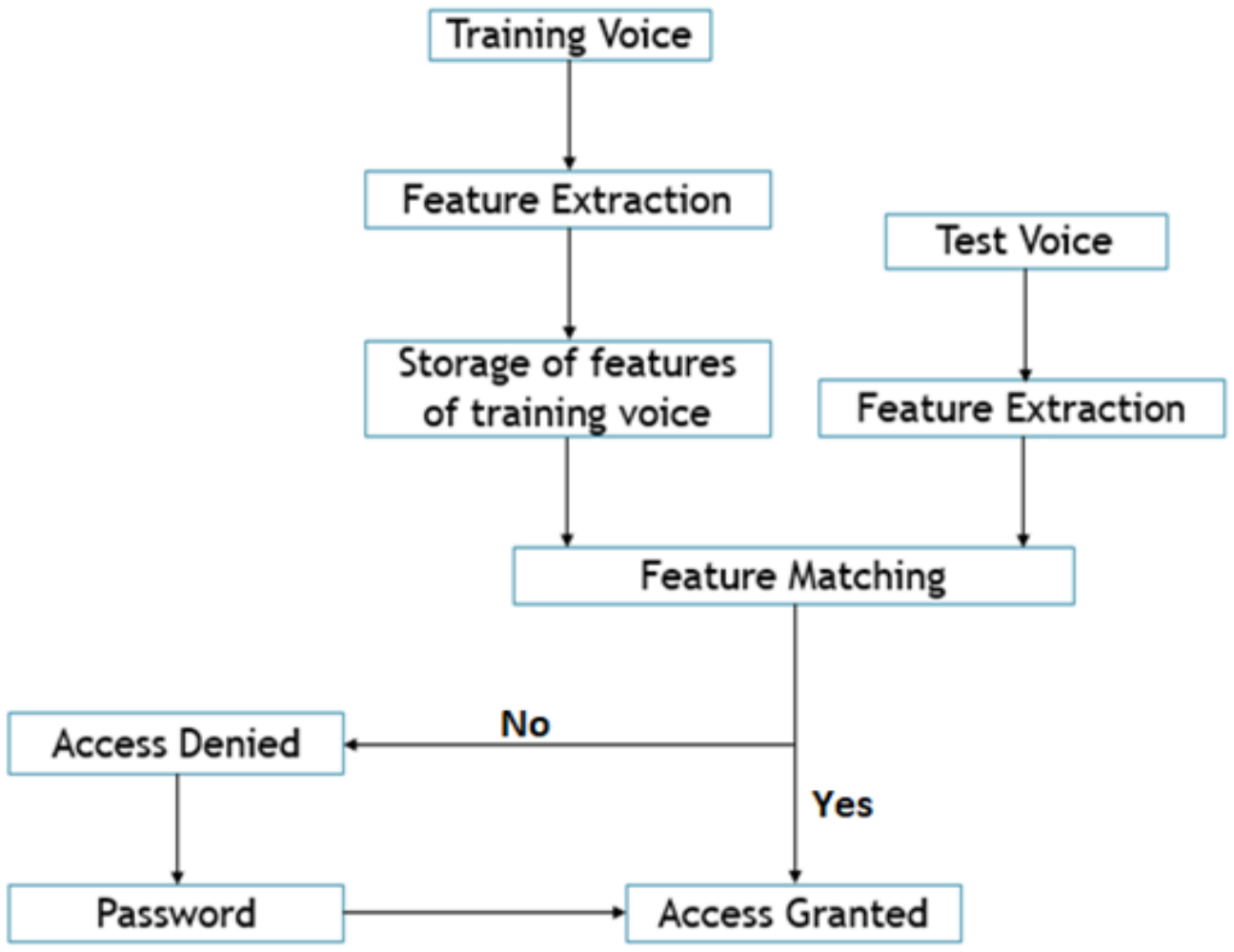

Fig. 1. Block diagram of the proposed system

### B. Signal Pre-processing and Feature Extraction

Before features can be extracted, it is often necessary to pre- process the data. Since the data is very noisy, the first step is to denoise the audio clips. Manual sorting was initially carried out to eliminate clips without speech or having large portion of leading or trailing noise. This was followed by applying wavelet denoising on the matrix of MATLAB vectors corresponding to selected 40 audio clips per each speaker. The weak signals were compensated through the pre-emphasis process, which entails the use of first order FIR filter. The filter is a direct implementation of the speech samples difference equation [11]

$$S'_n = S_n - kS_{n-1} \quad (1)$$

where $n = 1 \ldots N$, $N$ is the samples length and k is the pre-emphasis coefficient which should be in the range $0 \leq k < 1$, k value of 0.97 has been used. The pre-emphasized speech signal was subjected to short-time Fourier transform analysis with frame durations of 25ms and frame shifts of 10ms in order to partition the speech into frames, this is also known as frame blocking. The removal of disjointedness at the start and end of each frame was accomplished by applying Hamming window bandpass filtering of Equation 2, [2], [8] and [12],

$$w(n) = 0.54 - 0.46\cos\left(\frac{2\pi n}{N-1}\right), 0 \leq n \leq N-1 \quad (2)$$

where $n$ is the number of cepstral coefficients. Fast Fourier Transform (FFT) of the resulting signal produced the magnitude spectrum of the signal. Twenty triangular filters uniformly spaced on the Mel scale between lower and upper frequency (Hz) limits of the signal was designed and applied to the computed magnitude spectrum values to obtain filter-bank energies (FBEs).. The Mel scale was obtained as follows [1], [2] and [12]:

$$f_{Mel} = 2595 \times \log\left(1 + \frac{f_{Hz}}{700}\right) \quad (3)$$

The cepstral coefficients are calculated from log FBEs amplitude ($m_j$) using the discrete cosine transform as follows [11]:

$$C_i = \sqrt{\frac{2}{M}} \sum_{j=1}^{M} m_j \cos\left(\frac{\pi i}{M}(j-0.5)\right) \quad (4)$$

where M is number of filter bank channels and $i = 1 \ldots N$, $N$ is the number of cepstral coefficient. Final step applies sinusoidal lifter in Equation 5 to produce liftered MFCCs [13] and [14].

$$ceplifter = 1 + 0.5L\sin\left(\frac{\pi n}{L}\right), 0 \leq n \leq N-1 \quad (5)$$

where $N$ is the number of cepstral coefficients and $L$ is the cepstral lifter parameter. The stages involved in the feature extraction is shown in fig. 2.

### C. k-Nearest Neighbour Classification

kNN classifier determines the class of a data point by majority voting principle. The number of neighbors to be considered in voting is equivalent to the value of k, which was used to tune the algorithm for optimal performance. Prediction was done by taking the mean value of the k closest points. Euclidean distance method was used to find the distance between data points. 40 audio clips per speaker was used to train the algorithm, while 10 audio clips per speaker was used to test the algorithm ability to predict new audio clips of speakers.

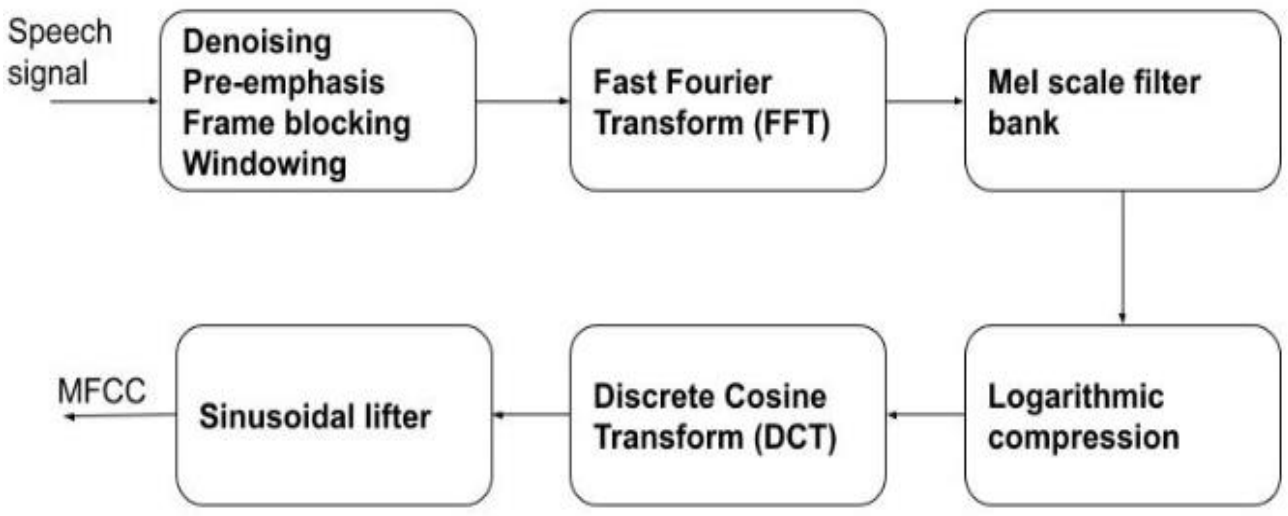

Fig. 2. Feature extraction stages

## IV. RESULTS

The effect of denoising the audio clips can be seen by comparing original sample audio clip with the corresponding denoise version as shown in Fig. 3. The wavelet was able to remove noise components successfully. Fig. 4 is the magnitude spectrum for a frame of 25ms of an audio clip, which shows that the lower frequencies have insignificant components. The designed triangular filter banks with 30 channels that is evenly spaced on the Mel scale can be seen in Fig. 5. The application of the filter bank on an audio frame will produce signal on the Mel scale similar to Fig. 6. The thirteen cepstral coefficients corresponding to a sample audio frame is shown in Fig. 7, while the lifter version is shown in Fig. 8. The first coefficient was plotted separately from the remaining 12 coefficients in each case. The discrimination of the coefficients was improved with the liftering. Table 1 shows the k-fold cross validation accuracy for different value of nearest neighbor. The highest accuracy obtained with manual tuning of k was 80%. The holdout cross-validation was able to yield 100% accuracy for 2 and 3 neighbors, but higher number of neighbors' accuracy is indeterminate.

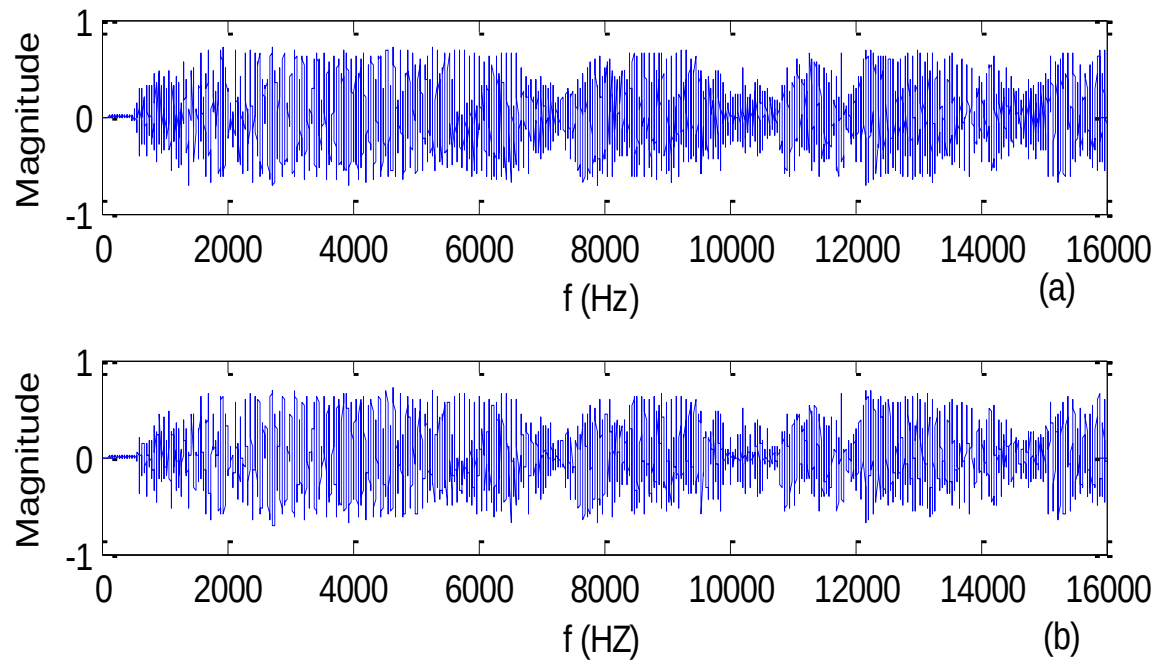

Fig. 3. Comparison of of (a) original signal with (b) denoised signal

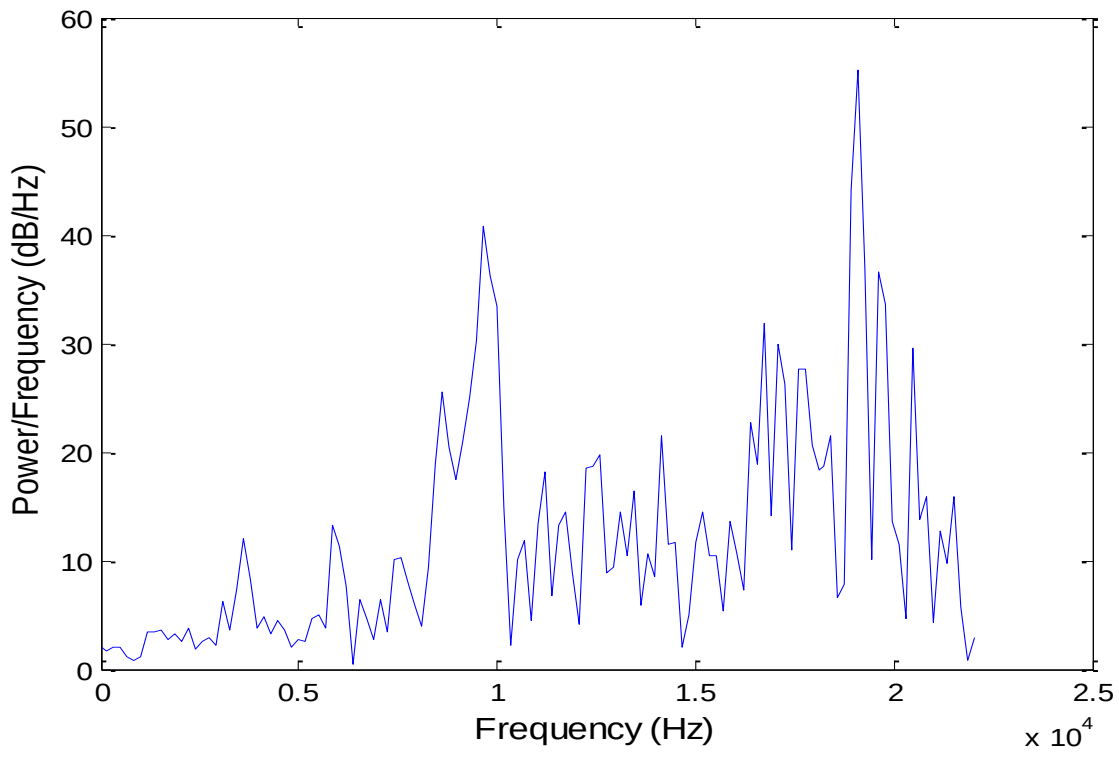

Fig. 4. Magnitude spectrum for 25ms clip

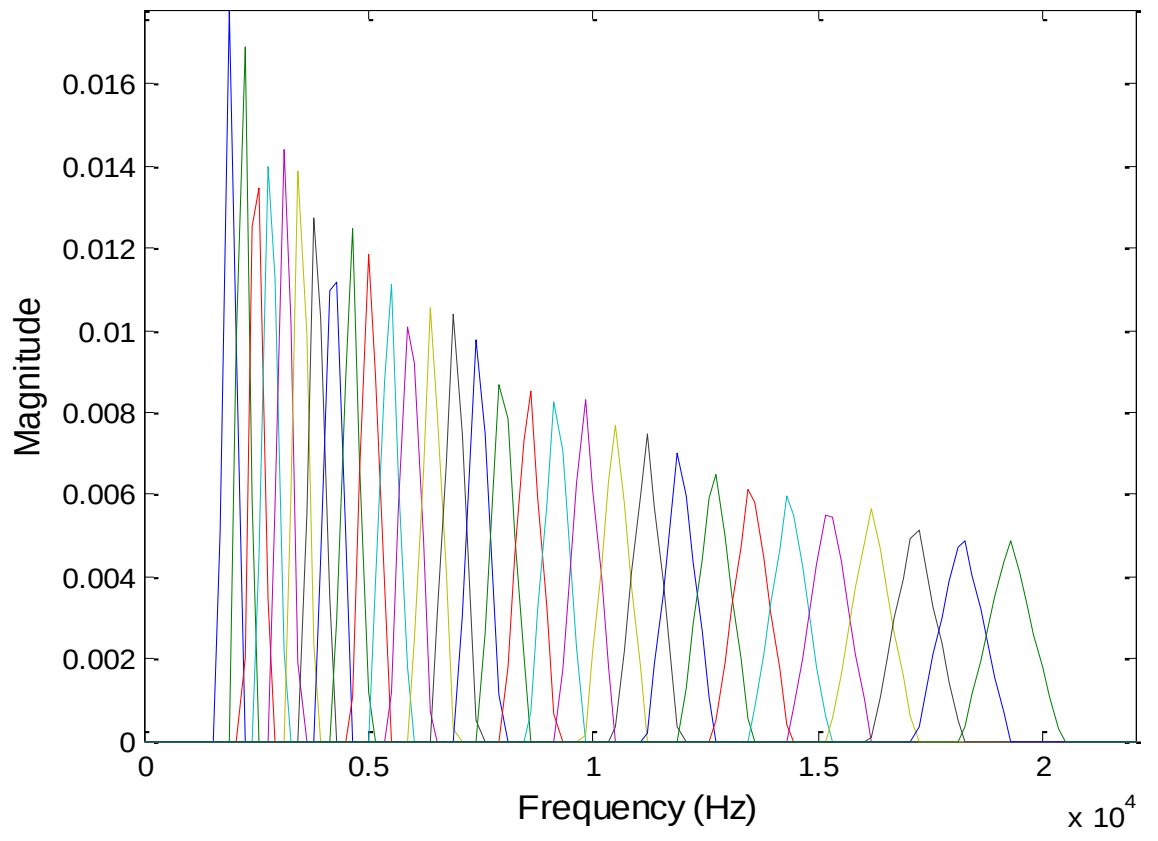

Fig. 5. 30-channel triangular filter banks

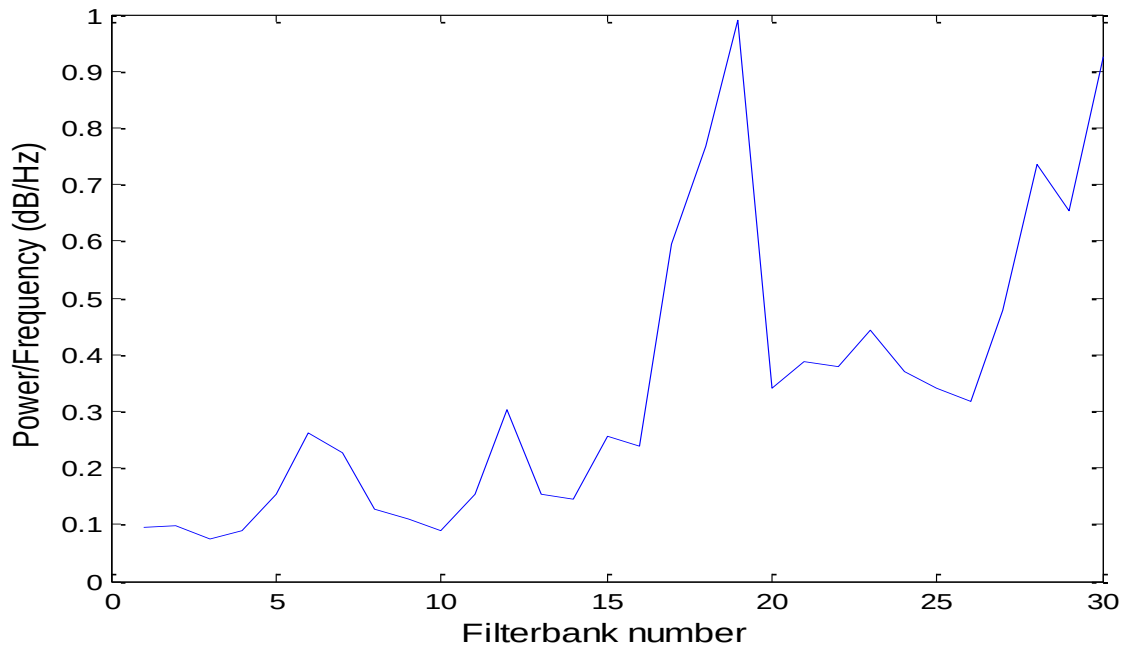

Fig. 6. Mel scale converted signal of a sample audio frame

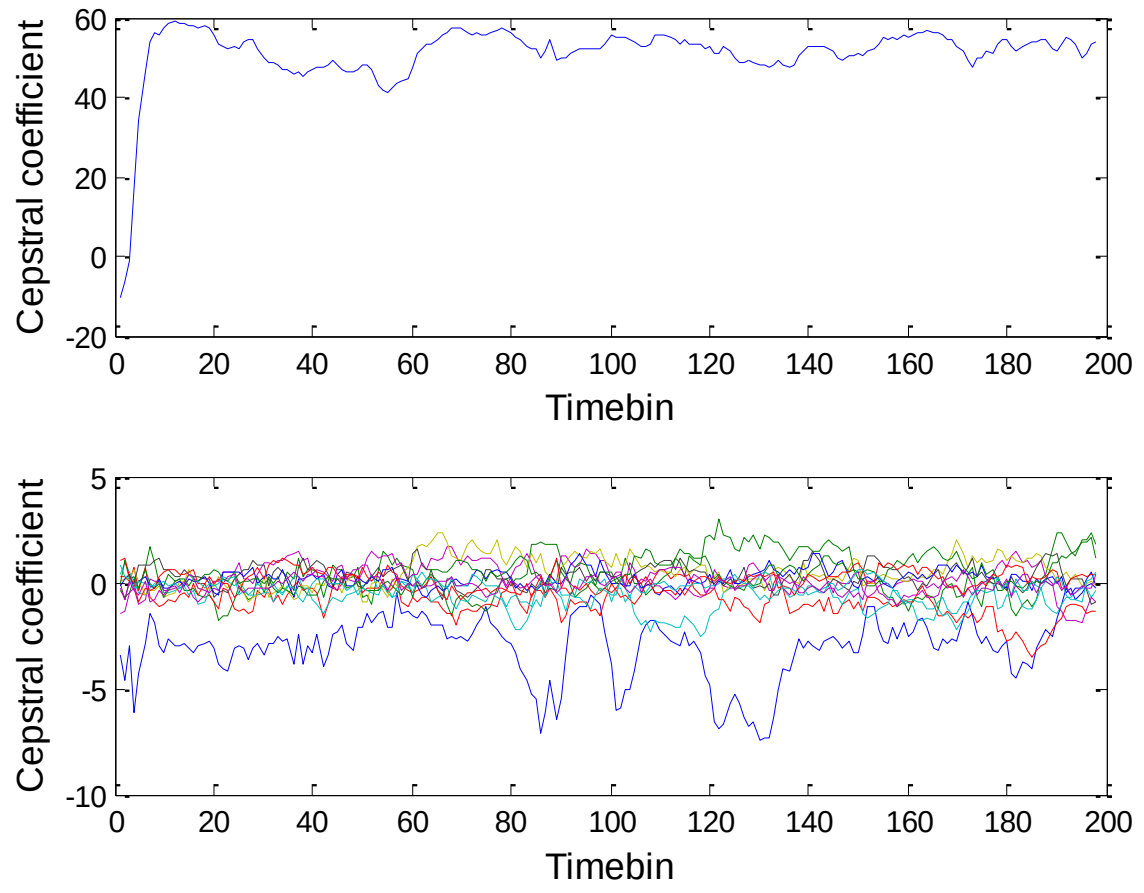

Fig. 7. First and 12 remaining cepstral components of sample frame without liftering

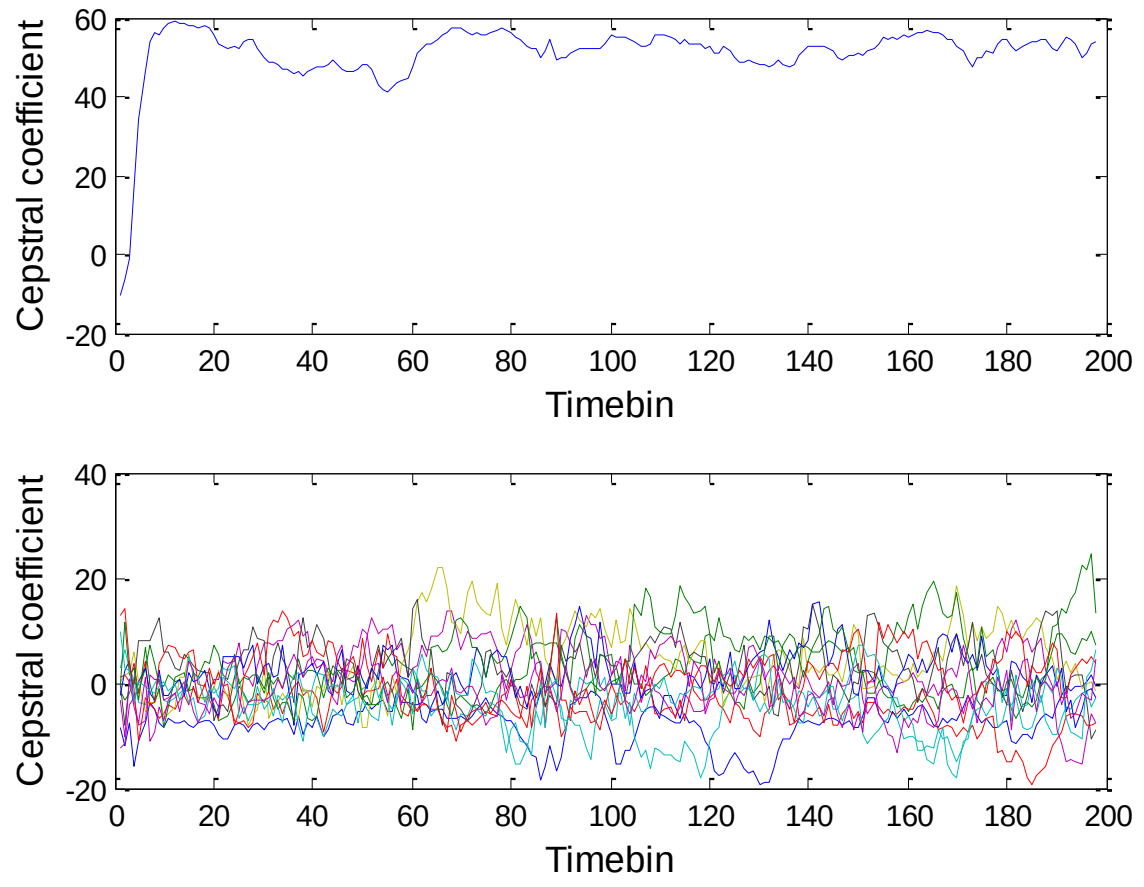

Fig. 8. First and 12 remaining cepstral components of sample frame with liftering

TABLE I. Variation of cross-validation accuracy with number of neighbors

| Number of neighbors | 2 | 3 | 4 | 5 | 6 | |
|---|---|---|---|---|---|---|
| K-fold accuracy | 20 | 60 | 80 | 60 | 60 | |
| Holdout accuracy | 100 | 100 | - | - | - | |

## V. Conclusion

Despite the pre-processing of the signal and manual tuning of the k-parameter, the algorithm could not achieve more than 80% k-fold cross validation accuracy. This is majorly due to the inherent poor learning performance of the algorithm. An algorithm with a better learning rate will be applied in the future. An improvement in accuracy is also possible if less noisy data is used or if it is text dependent. Exploring other feature extraction techniques may also produce an improvement. An audio-visual data can also be considered to improve the system.

## References

[1] M. T. S. Al-Kaltakchi, H. A. A. Taha, M. A. Shehab, and M. A. M. Abdullah, “Comparison of feature extraction and normalization methods for speaker recognition using grid-audiovisual database,” Indones. J. Electr. Eng. Comput. Sci., vol. 18, no. 2, pp. 782–789, 2020, doi: 10.11591/ijeecs.v18.i2.pp782-789.

[2] M. A. Nasr, M. Abd-Elnaby, A. S. El-Fishawy, S. El-Rabaie, and F. E. Abd El-Samie, “Speaker identification based on normalised pitch frquency and Mel Frequency Cepstral Coefficients,” Int. J. Speech Technol., vol. 21, no. 4, pp. 941–951, 2018, doi: 10.1007/s10772-018-9524-7.

[3] J. Ye, T. Kobayashi, M. Murakawa, and T. Higuchi, “Robust acoustic feature extraction for sound classification based on noise reduction,” in ICASSP, IEEE International Conference on Acoustics, Speech and Signal Processing - Proceedings, 2014, no. November, pp. 5944–5948, doi: 10.1109/ICASSP.2014.6854744.

[4] S. A. Alim and N. K. A. Rashid, “Some commonly used speech feature extraction algorithms,” in From Natural to Artificial Intelligence - Algorithms and Applications, R. Lopez-Ruiz, Ed. IntechOpen, 2018, pp. 2–19.

[5] Z. Bai and X. L. Zhang, “Speaker recognition based on deep learning: An overview,” Neural Networks, vol. 140. pp. 65–99, 2021, doi: 10.1016/j.neunet.2021.03.004.

[6] J. Lin, Y. Yumei, Z. Maosheng, C. Defeng, W. Chao, and W. Tonghan, “A multiscale chaotic feature extraction method for speaker r ecognition,” vol. 2020. Hindawi, p. 9 pages, 2020, doi: 10.1155/2020/8810901.

[7] R. T. Al-Hassani, D. C. Atilla, and Ç. Aydin, “Development of high accuracy classifier for the speaker recognition system,” Appl. Bionics Biomech., vol. 2021, no. 2, pp. 1–10, 2021, doi: 10.1155/2021/5559616.

[8] A. I. Siam, A. A. Elazm, N. A. El-Bahnasawy, G. M. El-Banby, and F. E. Abd El-Samie, “PPG-based human identification using Mel-frequency cepstral coefficients and neural networks,” Multimed. Tools Appl., vol. 80, no. 17, pp. 1–19, 2021, doi: 10.1007/s11042-021-10781-8.

[9] M. G. Pinto, M. Polignano, P. Lops, and G. Semeraro, “Emotions Understanding Model from Spoken Language Using Deep Neural Networks and Mel-Frequency Cepstral Coefficients,” in 2020 IEEE Conference on Evolving and Adaptive Intelligent Systems (EAIS), 2020, pp. 1–5, doi: 10.1109/EAIS48028.2020.9122698.

[10] S. Naeem et al., “Subspace Gaussian Mixture Model for Continuous Urdu Speech Recognition using Kaldi,” in 14th International Conference on Open Source Systems and Technologies (ICOSST), 2020, pp. 1–7, doi: 10.1109/ICOSST51357.2020.9333026.

[11] S. Young et al., The HTK Book (for HTK version 3.4). Cambridge: Cambridge University Engineering Department, 2006.

[12] A. Sithara, A. Thomas, and D. Mathew, “Study of MFCC and IHC feature extraction methods with probabilistic acoustic models for speaker biometric applications,” Procedia Comput. Sci., vol. 143, pp. 267–276, 2018, doi: 10.1016/j.procs.2018.10.395.

[13] K. K. Paliwal, “On the use of filter-bank energies as features for robust speech recognition,” in EUROSPEECH, 1999, pp. 641–644, doi: 10.1109/ISSPA.1999.815754.

[14] A. Benba, A. Jilbab, and A. Hammouch, “Detecting patients with Parkinson’s disease using mel frequency cepstral coefficients and support vector machines,” Int. J. Electr. Eng. Informatics, vol. 7, no. 2, pp. 297–307, 2015, doi: 10.15676/ijeei.2015.7.2.10.